\documentstyle[PASJadd,epsf]{PASJ95}

\makeatletter

\long\def\@makecaption#1#2{
 \vskip 10pt 
 \setbox\@tempboxa\hbox{#2}
 \ifdim \wd\@tempboxa >\hsize #2\par \else \hbox
to\hsize{\hfil\box\@tempboxa\hfil} 
 \fi}

\makeatother

\newcommand{\vol}{}
\newcommand{\no}{}
\newcommand{\lett}{}
\newcommand{\spage}{}
\newcommand{\rdate}{2000 December 26}
\newcommand{\adate}{2001 April 20}

\begin{document}

\title{The Radio-to-Submillimeter  Flux Density Ratio of Galaxies as a Measure 
of Redshift}

\author{T. N. {\sc Rengarajan} and Tsutomu T. {\sc Takeuchi}\\ 
{\it Division of Particle and Astrophysical Sciences, School of Science, 
Nagoya University}, \\
{\it Furo-cho, Chikusa-ku, Nagoya 464--8602} \\
{\it renga, takeuchi@u.phys.nagoya-u.ac.jp}
}

\abst{
We re-examine the technique of determining the redshifts of galaxies 
from the ratio of the submillimeter-to-radio continuum flux densities based on 
a recently published catalog of 850~$\mu$m sources. 
We derived the expected variation of this ratio as a function of redshift 
incorporating the expected average luminosity and the spectral energy 
distribution (SED) of dust emission and the radio continuum. 
We find that the existing data for most of the high redshift ($z \gtsim 1$)
sources correspond to our new calculation. Amongst the well-identified 
sources, there is none with an index significantly higher than predicted. 
Sources which have an index  lower than predicted are either within the 
error zone or much lower, the latter presumably having AGN-dominated 
radio-continuum emission. 
We find the median redshift to be $\sim 2$, which is consistent with 
that deduced by previous work ($\sim 3$) within the error. 
We also discuss the various systematic effects that can affect the 
accuracy of the redshift estimate. 
We examine other methods of redshift estimation, like photometric ratio 
in the submillimeter and locating the peak of the SED in the rest system 
of the objects. 
We conclude that while the various methods are helpful in identifying 
high-redshift objects and making a crude estimate of the 
redshift, they are not, at present, accurate enough for a detailed study 
of redshift distribution of the submillimeter galaxies.
}

\kword{galaxies: redshifts --- galaxies: starbursts --- infrared : galaxies 
--- radio continuum : galaxies}

\maketitle
\thispagestyle{headings}

\section{Introduction}   

It was first suggested by Carilli and Yun (1999; hereafter CY) that an 
estimate of the redshift of distant galaxies can be obtained from the ratio 
of the submillimeter (submm) to radio continuum flux densities. 
Following this work, several others have applied this technique to various 
samples (Smail et al.\ 2000; Barger et al.\ 2000). 
The former make use of a complete sample of 850~$\mu$m observations. 
This method has been used by several authors to obtain the redshift 
distribution of galaxies detected in the submm wavelengths 
(Smail et al.\ 2000; Barger et al.\ 1999; Bertoldi et al.\ 2000; 
Ivison et al.\ 2000). 
The basis for this method is the tight correlation between the far infrared 
(FIR) flux in a broad band of 40--120~$\mu$m (defined through the FIR 
parameter by Helou et al.\ 1988) and the radio continuum flux density (FD) 
at cm wavelengths (Condon 1992 and references therein). 
CY have discussed extensively the advantages of using these wavelengths. 
The steep nature of the submm SED and the fact that as the redshift increases,
the FD at the rest system submm wavelength increases sharply, while the 
corresponding radio FD decreases give rise to a fast change in the FD ratio 
with $z$. 
The accuracy of the method depends on the extent of the variation of SED 
from galaxy to galaxy.
Once the local value of the submm-to-radio ratio,
${\cal R}_{\rm sr} = S(353\; {\rm GHz})/S(1.49\; {\rm GHz})$, and 
the dust emission SED are known, the ${\cal R}_{\rm sr}$--$z$ behavior can be 
predicted. 

Most authors have used a modified blackbody spectrum with an emissivity 
dependence of $\epsilon(\nu) \propto \nu^{n}$ with $n = 1 \mbox{--} 1.5$ 
and the ${\cal R}_{\rm sr}$ value of nearby prototype galaxies, like 
M~82 or Arp 220, and other ultraluminous galaxies since a good sample of 
submm local galaxies was not available. 
Blain (1999a) and Carilli and Yun (2000a) have discussed the uncertainties 
associated with this method. 
Though they treated some important issues of real SEDs, many aspects are 
still overlooked and remain to be considered with regard to a large 
well-controlled sample of galaxies.

Recently, Dunne et al.\ (2000, hereafter DS-I) have published 
a catalog of 850 $\mu$m flux densities for a complete sample of 104 bright 
IRAS galaxies. 
This large sample can profitably be used to compute the expected variation of 
${\cal R}_{\rm sr}$ with $z$. 
In their erratum (Carilli, Yun 2000b) Carilli and Yun have also commented on 
the Dunne et al.\ data.
Therefore, we discuss this method further and compare available observations 
for well-identified high-$z$ galaxies with our computed curve. 
We also estimate the dispersion in the relationship and discuss the 
limitations of the method. 
Finally, we briefly comment on other methods proposed for a redshift 
determination, such as those using the submm photometric ratio 
(Hughes et al.\ 1998) and those locating the peak of the FIR emission 
(Blain 1999a, 1999b).

\section{Data and Analysis}  

\subsection{Submillimeter Data}

Dunne et al.\ (2000) tabulated the data for a complete sub-sample of 
IRAS galaxies selected from the Bright Galaxy Sample (Soifer et al.\ 1989). 
Besides other information, they tabulated the 850~$\mu$m flux densities 
measured using SCUBA and parameters of fits to the SED using their data 
and the IRAS 60 and 100 $\mu$m FDs. 
They assumed a modified blackbody spectrum characterized by dust temperature, 
$T_{\rm d}$ and a dust emissivity index, $n$. 
Though these values may not correctly represent the physical temperature, 
they give a reasonable measure of the SED at wavelengths longer than 60~$\mu$m. 
The mean and dispersion of $T_{\rm d}$ and $n$ are 35.6$\pm 4.9$~K and 
$1.3 \pm 0.2$, respectively. 
Using the parameter ${\rm FIR} = 1.26 \times 10^{-14} \times [2.58\, S(60) 
+ S(100)]$ and $H_0 = 75\;{\rm Mpc^{-1}km\,s^{-1}}$ the computed FIR 
luminosity, $L_{{\rm FIR}}$, ranges from 10$^{10}$ to $10^{12}\; L_\odot$. 
Radio-continuum FDs at 1.49~GHz are available for almost all DS-I sources
from Condon et al.\ (1990, 1996). 
Using these along with the DS-I 850~$\mu$m data, we find the mean and 
dispersion of log ${\cal R}_{\rm sr}$ to be 0.43 and 0.24, respectively. From 
a more careful look at the data, it is seen that $log {\cal R}_{\rm sr}$ is 
a decreasing function of $L_{\rm FIR}$, changing from 0.55 in the $\log
L_{{\rm FIR}}/L_\odot$ interval $10\mbox{--}10.3$ to 0.29 for the 
$11.2\mbox{--}12$ interval. 
This result arises mainly due to the correlation between ($T_{\rm d}, n$) 
and $L_{{\rm FIR}}$ affecting the ratio $S(353\;{\rm GHz})/{\rm FIR}$. 
We discuss the effect of this later. 
If a galaxy located at a redshift $z$ is observed at frequencies 
$\nu_{\rm s}^{\rm obs}$ and $\nu_{\rm r}^{\rm obs}$ at the Earth where the
subscripts s and r refer to submm and radio, respectively, the observed ratio
is related to the rest system ratio by the relation
\begin{eqnarray} 
  {\cal R}_{\rm sr} = \frac{S(\nu_{\rm s}^{\rm obs})}{
    S(\nu_{\rm r}^{\rm obs})} 
  = \frac{S[(1+z)\nu_{\rm s}^{\rm em}]}{S[(1+z)\nu_{\rm r}^{\rm em}]} \; .
\end{eqnarray}
If the radio SED is a power law of the form, $S(\nu_{\rm r}) \propto 
\nu_{\rm r}^{\alpha}$, the denominator can be replaced by 
$(1+z)^{\alpha} S(\nu_{\rm r}^{\rm em}$). 
For a subsequent discussion, we take $\nu_{\rm s} = 353$~GHz and 
$\nu_{\rm r} = 1.49$~GHz. 
For each source of DS-I, we compute $S[353(1+z)]$ using the fitted values of 
$T_{\rm d}$ and $n$ for that source and take the ratio to the observed 
1.49~GHz FD. 
The mean and dispersion of the ratio are then computed as a function of 
redshift. 
What about the effect of a luminosity--${\cal R}_{\rm sr}$ correlation? 
Though the submm-radio ratio for the local sample decreases by 
$\sim 0.25$~dex in the $\log L_{\rm FIR}$ range 10--12, the effect is diluted 
as $z$ increases, since one samples wavelengths closer to $100\;\mu$m
the corresponding decrease at redshifts of 1, 3, and 5 are only 0.15, 0.07, 
and 0.05, respectively, much less than the dispersions. 
These decreases are somewhat less than what one may expect from the SED 
variation as a function of FIR luminosity, since they are compensated by a 
decrease in radio emission due to self absorption for high-luminosity sources.
We consider this effect in greater detail while discussing the radio 
spectral index. 
It is well known that the apparent $850 \;\mu$m brightness of a source of 
given luminosity decreases slowly because the slope of the dust emission SED 
compensates for the distance (Blain, Longair, 1996). 
For a detection sensitivity of 5~mJy at 850~$\mu$m, the threshold value of 
$\log L_{\rm FIR}/L_\odot$ increases from 11.7 to 12 in the redshift interval 
0.5--6 [assuming the average observed ratio of S(353 GHz)/FIR and using 
$q_{0} = 0.5$]. 
On the other hand, for a given detection sensitivity of the radio continuum 
emission, the threshold FIR luminosity increases faster. 
For example, taking the detection sensitivity at 1.49~GHz to be 10~$\mu$Jy, 
we find that the calculated threshold luminosity  based on the 
well known FIR-radio correlation changes from 
$\log L_{\rm FIR}/L_\odot$ = 10.4 at $z = 0.5$ to 12.5 at $z = 5$. 
If we require detection at both 850~$\mu$m and 1.49~GHz, the higher of the 
above defines the threshold luminosity. 
Assuming the effective range of luminosity to be from the threshold to 
0.5~dex higher, the relevant logarithmic luminosity intervals for the range 
$z = 0.5\mbox{--}3$ and 3--6 are about 12--12.5~$L_\odot$ and 
12.5--13~$L_\odot$ respectively. 
{}From the data of DS-I, we find that the mean dust temperature in the $\log
L_{\rm FIR}/L_\odot$ interval of 11.2--12 is 39.5~K. 
Based on the observed $T_{\rm d}$--$L_{\rm FIR}$ correlation, the 
extrapolated temperatures are 42.4 and 44.6 K for the two luminosity 
intervals above 12. 
We, then, compute corrections to the 
$S[353(1+z)\,\mbox{GHz}]/S(1.49\,\mbox{GHz})$ ratio based on the changes 
in the ratios of $S[353(1+z)]/{\rm FIR}$ computed using the temperatures for 
the relevant luminosity interval and an emissivity index of 1.3. 
It may be noted that the application of the correction for the radio continuum 
SED transforms the above ratio to ${\cal R}_{\rm sr}$, the rest system 
submm-to-radio continuum FD ratio.
The corrected values are lower compared to the values determined from the DS-I 
data for the 11.2--12 interval. 
The corrections are 0.23, 0.2, and 0.15~dex at $z = 0.5$, 3, and 6, 
respectively.
It may be noted that the correction decreases for larger $z$ because one is 
sampling wavelengths closer to 100~$\mu$m.

\subsection{Radio Spectral Index}

In order to compute the expected radio continuum emission at frequencies 
higher than 1.49~GHz we need a knowledge of both the FD at $z \approx 0$ and 
the radio spectral index. 
CY assumed a spectral index of $-0.8$. 
In order to take into account the self absorption at frequencies close to 
1.49~GHz, Barger et al.\ (2000) estimated the FD at 1.49~GHz by 
extrapolating the observed FD for Arp 220 at a high frequency of 8.4~GHz 
using a spectral index of $-0.8$ and using the same index for computing FD at 
intermediate frequencies. 
It is true that the observed 1.49~GHz FD is reduced due to self-absorption 
effect, especially for luminous sources, thus increasing the value of 
${\cal R}_{\rm sr}$. 
However, it should be noted that when a distant source is observed, what we 
measure is not the intrinsic flux, but the emergent flux including the effect 
of self absorption. 
Therefore, it is more appropriate to use the observed 1.49~GHz values and 
make a correction to the expected behavior while computing the submm--radio 
spectral index as a function of $z$. 
Using the available radio data (1.49 GHz: Condon et al.\ 1990; 4.85 GHz: 
Condon \& Broderick, 1991; Condon et al.\ 1995), we find the mean and 
dispersion of the 4.85--1.49 GHz spectral index for the available DS-I 
sources to be $-0.65$ and 0.26 respectively. 
For 21 DS-I sources for which 8.44~GHz data are available 
(Condon et al.\ 1991) the 1.49--8.44~GHz spectral index is $-0.47$. 
These galaxies have $\log L_{\rm FIR}/L_\odot$ greater than 11. 
Stine (1992) finds, for a sample of 23 Markarian galaxies, the 1.4--5~GHz 
spectral index to be $-0.64$. From 
these data it is clear that the absorption effect at low frequencies is 
present. 
Condon et al.\ (1991) observed a sample of 40 ultraluminous infrared galaxies, 
and found that the FIR-to-1.49~GHz radio continuum ratio for luminous 
($\log L_{\rm FIR}/L_\odot > 11.4$) sources containing compact radio 
cores is 0.25 dex higher than that for the rest of the galaxies, which have 
the standard ratio. 
When they correct for the self absorption at 1.49~GHz using the 8.44~GHz FD 
and assuming the real radio spectral index to be $-0.7$, the corrected mean 
ratio is normal. 
If we assume that the self absorption effect is confined to $\nu < 5$~GHz, 
the spectral index in the 1.49--5~GHz range is then $\approx -0.3$. 
In the DS-I data we also find the effect of self absorption in 
the FIR to radio continuum ratio. 
The 4.7--1.4 GHz spectral index of Arp~220 (Anantharamaiah et al.\ 2000) 
is $-0.3$. 
Taking all the above into account, a reasonable approximation for the radio 
spectral index would be $-0.4$ in the 1.49--5 GHz range and $-0.7$ for 
frequencies higher than 5~GHz and to scale FDs from the value at 1.49~GHz. 
Using this approximation and the method enumerated earlier for computing the 
$S[353(1+z)\; {\rm GHz}]/S(1.49\; {\rm GHz})$ ratio, we calculate 
${\cal R}_{\rm sr}$ as a function of $z$ using equation~1.
 
\subsection{Relationship between Redshift and Submm-Radio Flux Ratio}

The 353--1.49 GHz spectral index computed from this is shown in figure~1 as 
a solid line. 
For estimating the dispersion, we use the computed dispersion in the 
$S[353(1+z)]/S(1.49)$ ratio and also assume $\delta\alpha_{\rm r} = 0.2$. 
The dispersion of the spectral index is between 0.1 and 0.11. 
The upper and lower envelopes are shown as dashed lines in figure~1. 
Our values of the ratio $S[353(1+z)]/S(1.49)$ agree within 0.1~dex with those 
obtained by Barger et al.\ (2000) based on the SED and FD of Arp~220; 
for $z < 1$, our values are lower and for higher $z$, larger. 
However, after the application of radio spectral index effects, our 353--1.49 
spectral index values are about 0.05 lower throughout. 
At $z \approx 0$, our index is higher reflecting the free--free absorption 
effect. 
It may be noted that our result is not based on a single prototype, 
but on the data of over 20 galaxies with $\log L_{\rm FIR}/L_\odot > 11.2$.

\section{Discussion}  

\subsection{Comparison with Observations}

\begin{figure}[t]
\epsfxsize=7.0cm
\centerline{\epsfbox{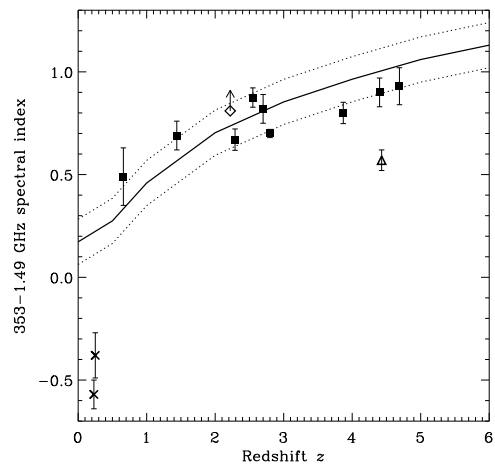}}
\caption{Fig.\ 1\ ---
  The 353 GHz-to-1.49 GHz spectral index, $\gamma$, plotted as a function of 
  galaxy redshift, $z$. 
  The solid line is the computed line and the two dashed lines are the 
  envelopes of dispersion (see text for details). 
  Symbols: filled squares -- both $\gamma$ and $z$ definite; 
  open diamonds -- $z$ certain, but radio FD upper limit; 
  crosses -- cD galaxies;
  open triangle -- a dusty QSO BRI~$0952 - 0115$.
  For references to objects and data, refer to subsection 3.1 of the main 
  text.}
\end{figure}


In figure~1 we also show the data points of individual high-$z$ sources 
compiled from available publications. We have included only well-identified 
sources that have definite redshift measurements from the near-infrared or CO 
spectral lines. 
The objects with the values of redshift in parentheses and 
references to them are: 
IRAS~F$10214+0724$ (2.29) (Rowan-Robinson et al.\ 1993); 
HR~10 (1.44) (Cimatti et al.\ 1998; Dey et al.\ 1999);
CFRS 14F (0.66) (Lilly et al.\ 1999); 
BR~$1202-0725$ (4.69) (Isaak et al.\ 1994); 
SMM~J$14011+0252$ (2.55) (Ivison et al.\ 2000); 
SMM~J$02399-0136$ (2.8) (Ivison et al.\ 1998); 
APM~$08279+5225$ (3.87) (Lewis et al.\ 1998); 
SMM~J$21536+1741$ (0.23), SMM~J$14010+0252$ (0.25) ( Edge et al.\ 1999);
SMM~J$14010+0253$ (2.22) (Ivison et al.\ 2000); 
LBQS~$1230-1627$B (2.7), BRI~$1335-0417$ (4.4), BRI~$0952-0115$ (4.43) (Yun et 
al. \ 2000).
We have included CFRS 14F, since Lilly et al.\ (1999) consider it to be their 
most secure identification and SMM~J$14010+0253$ based on the spectral line 
identification of Ivison et al.\ (2000). The 1.4~GHz FD for CFRS 14F has been 
extrapolated from the 5~GHz FD assuming a spectral index of $-0.7$.  
For HR~10, we have used the FDs listed by Dey et al.\ (1999); 
the 1.4~GHz FD has been extrapolated from the 3.6 cm measurement. 
It may be noted that for the three dusty QSOs of Yun et al.\ (2000) 
the index is an extrapolation from the 240~GHz measurement assuming 
a 350--250 GHz spectral index of $-3.5$. 
For their fourth source, BR~$1202-0725$, we used the 350~GHz flux density 
measured by Isaak et al.\ (1994). 
The two Edge et al.\ (1999) sources are cD galaxies in the centers of rich 
clusters of galaxies, A~1835 and A~2390. 

The curve shown in figure~1 is representative of the mean for a sample of 
galaxies at different redshifts. 
Individual galaxies may deviate from the curve. 
An inspection of figure~1 shows that the observed data points populate the 
region delineated by our calculations. 
The exceptions are the two cD galaxies and the dusty QSO, BRI~$0952-0115$,
which are well below the predicted range. 
These sources must be dominated by AGN radio emission. 
It is interesting that for this sample of well-identified sources, 
there is none occurring above the predicted error zone. 
We also note that the AGN-dominated sources seem to be clearly separated 
from the starburst galaxies. 

What are the types of sources that may populate regions above the predicted 
zone? 
The submm SED may be significantly affected by the presence of cold dust. 
{}From ISO observations of a sample of `active' and `inactive' galaxies 
covering a luminosity range of $\log L_{\rm FIR}/L_\odot = 10 \mbox{--} 11.2$,
Siebenmorgen, Kr\"{u}gel and Chini (1999) conclude that while the SED of the 
former sample is well fitted by a modified blackbody spectrum with a single 
temperature of $\sim 30\mbox{--}32$~K, the latter `inactive' sample requires 
the presence of an additional colder component ($T = 11 \mbox{--}16$~K). 
These galaxies have a higher submm-to-FIR flux ratio, and will yield a 
353~GHz--1.4~GHz index higher by up to 0.3 as compared to our calculation. 
From an analysis of their data, we find that the redshift estimate will be 
most affected at wavelengths close to 850~$\mu$m ($z < 1 \mbox{--} 1.5$)
for such galaxies. 
Such effects may explain the discrepancy between the redshift estimates using 
the submm--FIR photometric ratio and the submm-radio index for FIRBACK sources 
of Scott et al.\ (2000). 
`Inactive' galaxies with cold dust may be a significant fraction of 
the low-$z$ population, and hence an estimate of the redshift from the 
submm--radio index could distort the picture of the star-formation rate 
at low redshifts. 
On the other hand, it is unlikely that such sources are present at high 
redshifts, since we do not expect very luminous `inactive' galaxies. 

It is of interest to see what is the median redshift of the complete 
Smail et al.\ (2000) sample of galaxies. 
If we use the spectroscopic redshifts for 5 galaxies with reliable 
identification and estimate the redshift for the remaining 11 galaxies from 
our submm--radio index curve, we find the median redshift to be $\sim 2.1$. 
In the above, if the estimate is a lower limit, the source is assumed to be 
at that redshift value. 
If, on the other hand, we use only the estimated redshift from the 
submm-radio index and exclude the two cD galaxies, the median redshift is 
$\sim 2.2$. 
These values are similar to those deduced by Smail et al.\ (2000) within the
error of $\Delta z \sim 1$ at these redshifts.
  
What are the systematic effects one would expect at higher redshifts? 
As remarked earlier, higher $z$ sources end to be more luminous, 
have a steeper FIR--submm SED, and hence, lower 353--1.49 GHz spectral index. 
If sources with high temperature ($T > 50$~K with $n =1.3$) are present, 
they would populate regions lower than the predicted curve. 
However, there are other effects which may compensate this. 
If the free--free absorption is high, the radio continuum FD, even at higher 
frequencies, could be lower, thus increasing the observed 353--1.49 GHz index. 
Another effect could arise if the emission at 100~$\mu$m is optically thick. 
Solomon et al.\ (1997), who observed a sample of very high luminosity 
sources out to redshift of 0.3 in the CO line, conclude that $\tau_{100}$ is 
high for these sources. 
If the 100 $\mu$m optical depth is large, say $\approx 1$, the emission can 
still be optically thin at submm wavelengths, thus increasing the submm--FIR 
ratio. 
For an SED characterized by an emissivity index, $n = 1$, and temperature 
range of $45\mbox{--}55$~K, the change in the logarithmic ratio of 
$S(\nu)/{\rm FIR}$ for the cases $\tau_{100} = 1$ and $\tau_{100} \ll 1$ 
are increases of 0.17 to 0.24 as the wavelength changes from 200 to 
600~$\mu$m. 
These are about the same as the decrease in the individual ratio for 
$\tau_{100} \ll 1$ while the temperature changes from 45 to 55~K. 
The presence of an AGN core would also depress the observed ratio. 
This aspect has been discussed in detail by CY.

\subsection{Accuracy of Redshift Estimate}

If we want to estimate the redshift from the value of the ratio of 
submm-to-radio continuum FDs, the uncertainty in redshift is given by 
the intersection of a horizontal line at that value with the dispersion 
envelopes shown in figure~1. 
We, then, find the uncertainties in the redshift estimate at $z = 1, 2, 3, 4$, 
and 5 to be $\pm 0.33, (+0.7,-0.5), (+0.9,-0.8), (+1.1,-1)$, and $\pm 1.2$, 
respectively. 
These uncertainties do not include the systematic effects discussed in the 
previous section. 
At $z \ltsim 2$, the redshift estimate would be reasonably accurate. 
However, the changes in SED of the source affect the result more than in the 
high-$z$ region, since the sampled wavelengths are closer to $850\;\mu$m. 
If there is a cold dust component present, the observed spectral index would 
be higher than the predicted value, and less if the temperature is high. 
The effects of the SED variation are less for high $z$, but the flattening 
gradient gives rise to larger errors. 
An additional problem is the effect of the large $\tau_{100}$ if the sources 
are very luminous and compact. 
At $z\gtsim 3$, the lower limit is perhaps the best result. 
For $z> 7$, the sensitivity of the technique would be very poor, since the 
sampled wavelengths are in the flat peak region of the SED. 
In summary, the 353--1.49~GHz ratio can lead to a crude estimate of a redshift 
and reasonable lower limits. 
Hence at present, the study of the distribution of redshifts from just this 
ratio is premature. 
The situation will perhaps improve once a larger sample of well-identified 
galaxies is available and with better knowledge of the systematic effects.

\subsection{Photometric Redshift}

Hughes et al.\ (1998) have suggested that the redshift can be estimated from 
the ratio of FDs in two or more submm bands. For this technique to work, at 
least one rest system wavelength should be \ltsim 200~$\mu$m, away from the 
single power-law region. 
Scott et al. \ (2000) have also used the FIR-to-submm FD ratio for their 
FIRBACK sample.
The uncertainties arising from variations in SED will be similar to the 
submm--radio index method. 
Unlike the latter method that uses only one rest wavelength sampling, the 
steep part of the SED, the requirement of measurements at two bands, at least 
one rest wavelength away from this steep region, reduces the sensitivity 
of the photometric method. 
For rest wavelengths $> 1000 \;\mu$m, significant contamination from the 
thermal free--free emission may be present. 
This method can be a useful supplement to the submm--radio index method,
especially for a redshift range of 1--3.

\subsection{Redshift from the Peak of SED}

Blain (1999a, 1999b) has suggested that the redshift could be estimated by 
locating the peak of the dust emission SED. 
This technique works if all sources are characterized by a constant 
temperature. 
Uncertainty in redshift is directly proportional to the changes in the peak 
wavelength that depends on $T$, the emissivity index, and the optical depth 
near the peak. 
Another factor affecting the redshift estimate is the accuracy with which peak 
frequency can be located.  
Apart from the observational constraints arising from the availability of 
suitable bands for measurement, uncertainty is also introduced by the 
flatness of the SED in the peak region. 
We estimate that the resulting error in $z$ is $\sim 0.3$ at $z = 1$ and 
$\sim 0.9$ at $z = 5$. 
The uncertainty in SED parameters would further increase the errors. 
If multiband photometry on both sides of the peak are available, 
a template SED can be fitted to get better accuracy. 

\section{Conclusions}    

We have re-examined the method for estimating the redshifts of galaxies 
using the submm--radio spectral index. 
For this purpose, we made use of the properties of the 850~$\mu$m 
sample of galaxies observed by Dunne et al.\ (2000). 
We compute the variation of the 350--1.4 GHz spectral index as a function of 
$z$ taking into account the effects of luminosity on the SED and free--free 
absorption effects on the radio emission. 
We find that all well-identified sources have their spectral indices either 
within the error zone, or much lower. 
The latter, presumably radio-loud AGNs, are well separated from the rest. 
We also find the median redshift to be $\sim 2$, which is consistent with
that deduced by Smail et al.\ (2000) within an error of $\Delta z \sim 1$
at these redshifts. 

\vspace{1pc}\par
TNR thanks the Japan Society for the Promotion of Science for the award of 
an Invitation Fellowship and Dr.\ Hiroshi Shibai for his hospitality.

\section*{References}
\small

\re
Anantharamaiah, K. R., Viallefond, F., Mohan, N. R., Goss, W, M., \&
Zhao, J. H.\ 2000, ApJ, 537, 613  
\re
Barger, A. J., Cowie, L. L., \& Richards, E. A.\ 2000, AJ, 119, 2092
\re
Bertoldi, F., Carilli, C. L., Menten, K. M., Owen, F., Dey, A., Gueth, F., 
Graham, J. R., Kreysa, E. et al.\ 2000,\ A\&A, 360, 92 
\re
Blain, A. W. 1999a,\ MNRAS, 309, 955  
\re
Blain, A. W. 1999b, in Weyman, R.J., Storrie-Lombardi L., Sawicki M., Brunner
R. ed., ASP Conf. Ser. 191, Photometric Redshifts and High Redshift Galaxies,
255
\re
Blain, A. W., \& Longair, M. S. 1996, MNRAS, 279, 847
\re
Carilli, C. L., \& Yun, M. S. 1999,\ ApJ, 513, L13 (CY)  
\re
Carilli, C. L., \& Yun, M. S. 2000a,\ ApJ, 530, 618
\re
Carilli, C. L., \& Yun, M. S. 2000b,\ ApJ, 539, 1024 (erratum)
\re
Cimatti, A., Andreani, P., Rottgering, H., \& Tilnaus, R.\ 1998, 
Nature, 392, 895  
\re
Condon, J. J.\ 1992, ARA\&A, 30, 575  
\re
Condon, J. J., Anderson, E., \& Broderick, J. J.\ 1995, AJ, 109, 2318  
\re
Condon, J. J., \& Broderick, J. J.\ 1991, AJ, 102, 1663  
\re
Condon, J. J., Helou, G., Sanders, D. B., \& Soifer, B. T.\ 1990, ApJS, 73, 
359  
\re
Condon, J. J., Helou, G., Sanders, D. B., \& Soifer, B. T.\ 1996, ApJS, 103, 
81  
\re
Condon, J. J., Huang, Z.-P., Yin, Q. F., \& Thuan, T. X.\ 1991, ApJ, 378, 65  
\re
Dey, A., Graham, J. R., Ivison, R. J., Smail. I., Wright, G. S., \& 
Liu, M. C.\ 1999, ApJ, 519, 610  
\re
Dunne, L., Eales, S., Edmunds, M., Ivison, R., Alexander, P., \& 
Clements, D. L.\ 2000, MNRAS, 315, 115 (DS-I)  
\re
Edge, A. C., Ivision, R. J., Smail, I., Blain, A. W., \& Kneib, J.-P. 1999, 
MNRAS, 306, 599
\re
Helou, G., Khan, I. R., Malek, L., \& Boehmer, L.\ 1988, ApJS, 68, 151  
\re
Hughes, D. H., Serjeant, S., Dunlop, J., Rowan-Robinson, M., Blain, A., 
Mann, R. G., Ivison, R., Peacock, J., et al.\ 1998, Nature, 394, 241  
\re
Isaak, K. G., McMahon, R. G., Hills, R. E., \& Withington, S.\ 1994, MNRAS, 
269, L28  
\re
Ivison, R. J., Smail, I., Barger, A. J., Kneib, J.-P., Blain, A. W., 
Owen, F. N., Kerr, T. H., \& Cowie, L. L.\ 2000, MNRAS, 315, 209  
\re 
Ivison, R. J., Smail, I., Le Borgne, J. -F., Blain, A. W., Kneib, J.-P., 
Bezecourt, J., Kerr, T. H., \& Davies, J. K.\ 1998, MNRAS, 298, 583  
\re
Lewis, G. F. Chapman, S. C., Ibata, R. A., Irwin, M. J., \& Totten, 
E. J.\ 1998, ApJ, 505, L1  
\re
Lilly, S.J., Eales, S. A., Gear, W. K. P., Hammer, F., Le F\`{e}vre, O., 
Crampton, D., Bond, J. R., \& Dunne, L.\ 1999, ApJ, 518, 641  
\re
Rowan-Robinson, M., Eftasthiou, A., Lawrence, A., Oliver, S., Taylor, A., 
Broadhurst, T. J., McMahon, R. G., Benn C. R., et al.\ 1993, MNRAS, 261, 513  
\re
Scott, D., Lagache, G., Borys, C., Chapman, S. C., Halpern, M., Sajina, A., 
Ciliegi, P., Clements, D. L., et al.\ 2000, A\&A, 357, L5
\re
Siebenmorgen, R., Kr\"{u}gel, E., \& Chini, R.\ 1999, A\&A, 351, 495  
\re
Smail, I., Ivison, R. J., Owen, F. N., Blain, A. W., \& Kneib, J.-P.\ 2000, 
ApJ, 528, 612  
\re
Soifer, B. T., Boehmer, L., Neugebauer, G., \& Sanders, D. B.\ 1989, 
AJ, 98, 766 
\re
Solomon, P. M., Downes, D., Radford, S. J. E., \& Barrett, J. W.\ 1997, 
ApJ, 478, 144
\re
Stine, P. C.\ 1992, ApJS, 81, 49  
\re
Yun, M .S., Carilli, C. L., Kawabe, R., Tutui, Y., Kohno, K., \& Ohta, K. 2000,
ApJ, 528, 171

\label{last}

\end{document}